# Independent Manipulation of Electric and Thermal Fields with Bilayer Structure

Chuwen Lan[1,2], Xianglong Yu[1], Lingling Wu[1], Bo Li[2]*, Ji Zhou[1]*

[1]State Key Laboratory of New Ceramics and Fine Processing, School of Materials Science and Engineering, Tsinghua University, Beijing 100084, China

[2]Advanced Materials Institute, Shenzhen Graduate School, Tsinghua University, Shenzhen, China

*Corresponding author: zhouji@mail.tsinghua.edu.cn, libo@mail.tsinghua.edu.cn,

**Abstract:**

**Recently, increasing attention has been focused on the employment of transformation and metamaterial for manipulation of various physical fields, which requires complicated configuration and usually limits in single field. Here, for the first time, we propose and experimentally demonstrated bilayer structure to achieve simultaneously independent manipulation of multi-physics field (*dc* electric fields and thermal) by directly solving the *dc* electric/ thermal field equations. This structure is composed of two layers: the outer layer is made of isotropic and homogeneous material, while the inner layer is fan-shape layer. Since it is not based on TO, it can be readily experimentally fabricated with naturally occurring materials. Experimentally, we has designed, fabricated and characterized two structures simultaneously behaving as *dc* electric cloak/ thermal concentrator and *dc* electric concentrator/ thermal cloak, respectively. The simulation results agree well with the experiment ones, thus confirming the feasibility of our methodology. This job provides a novel avenue to manipulation of multiphysics fields, thus might find potential applications in various areas.**

## Introduction

Seeking the manipulation of physical fields in a desired manner is a dream of physicists and engineers. To achieve this goal, materials or structures are usually employed. Yet, they are traditionally aiming at manipulation of single field or realization of single-target applications. Taking the integrated circuit as an example, the conductors, semiconductors, as well as junction structures are utilized to transfer or shape the currents, while the radiator is employed to dissipate heat. In fact, the multi-physics field is much common and inevitable in many cases. Realization of dependent or independent manipulation of multi-physics field using the single device would create much more freedom for the design of new systems and might enable new potential applications.

Recently, metamaterial, also called artificial atoms, is attracting much attention and research interest due to its novel way for manipulation of various physical fields, like electromagnetics (EM), acoustic, thermal fields, etc. Compared to traditional materials, metamaterial has greater flexibility since its properties are not relying on what materials it is made of, but are instead depended on the spatial arrangement and geometry of constituent inclusions. Motived by the fantastic properties like negative refraction index [1], perfect absorber [2], invisibility cloak [3-6], etc, it has emerged as a hot topic in research of physics and material science. Another important motivation is transformation optics (TO), which is based on the physical field invariance under the transformation of coordinates. With the combination of metamaterial, the TO provides a powerful tool to realize the manipulation of the physical field in a desired way, which has not only been applied to wave fields like electromagnetics [3-6], acoustic [7], elastodynamics [8], ect, but also to the scalar fields such as magnetostatic [9,10], $dc$ current [11,12], thermal field [13-18] and mass diffusion field [19-21]. However, the previous works are limited to the manipulation of single field. Just recently, an interesting work is reported on the theoretical analysis of independent manipulation of thermal and electric current based on TO [22]. By introducing the concept of bifunctional metamaterial, a shell with thousands of thermal and electric elements was carefully designed to behave as a thermal concentrator and electrical invisibility cloak. However, from the view of practical application, the use of metamaterial and TO may suffer many practical problems. For example, the TO-based concentrator and invisibility cloak usually required anisotropic, inhomogeneous, and even extremely parameters which can only be achieved by complex metamaterial structure, thus greatly complicating the fabrication, especially for the 3D case and scaling to nano size.

In this paper, we propose that bilayer structure can be employed to achieve this goal. We further provide, for the first time to our knowledge, the experimental demonstration of independent manipulation of thermal and electric fields simultaneously. Instead of utilizing TO method, we introduce this structure by directly solve the *dc* electric/ thermal field equations, thus avoiding the use of complex metamaterial structure. Using naturally occurring materials, we fabricated two bilayer shells which can behave as electrical cloak / thermal concentrator and electrical concentrator/ thermal cloak, respectively. The experimental results are in great agreement with the simulated ones, thus confirming the feasibility of our methodology.

## 2. Thermal Concentrator and *dc* Electric Cloak

The bilayer structure for independent manipulation of thermal and *dc* electric fields is schematically shown in Figure 1(a), where the heat flows from the high temperature $T_1$ to the low temperature $T_2$, meanwhile, the current flows from the high potential $V_1$ to the low potential $V_2$. The space is divided into three regions: the interior ($r < a$), the bilayer shell with inner layer ($a < r < b$) and outer layer ($b < r < c$), and the exterior ($r > c$). We first consider that the bilayer structure has the capacity of concentrating thermal flux into the exterior and cloaking such region from the electric current, namely thermal concentrator and *dc* electric field cloak. The corresponding schematic diagrams are shown in Figure 1(b) and Figure 1(c), respectively. Here, we assume that the thermal conductivity and electric conductivity of the background material are $\kappa_b$ and $\sigma_b$, while the ones for inner layer and outer layer of the bilayer structure are $\kappa_1$, $\sigma_1$ and $\kappa_2$, $\sigma_2$, respectively. The thermal field and electric current field here can be described by $\nabla \kappa (\nabla T) = 0$ and $\nabla \sigma (\nabla \phi) = 0$, respectively. Since the external fields, both for thermal field and electric current field, should be undisturbed, one can directly solve these two equations and obtain that [12, 16, 18]:

$$\frac{c^2}{b^2} = \frac{(\kappa_2 - \kappa_1)(\kappa_2 + \kappa_b)}{(\kappa_2 + \kappa_1)(\kappa_2 - \kappa_b)} \qquad (1)$$

$$\frac{c^2}{b^2} = \frac{(\sigma_2 - \sigma_1)(\sigma_2 + \sigma_b)}{(\sigma_2 + \sigma_1)(\sigma_2 - \sigma_b)} \tag{2}$$

To realize a *dc* electric field cloak, the electric conductivity of the inner layer is set as 0 to keep the interior from the current, hence, the required electric conductivity is

$$\frac{c^2}{b^2} = \frac{\sigma_2 + \sigma_b}{\sigma_2 - \sigma_b} \tag{3}$$

To achieve a thermal concentrator, we assume that the thermal conductivity of the outer layer is homogeneous and anisotropic, while the one for the inner layer is homogeneous but anisotropic with radial component $\kappa_r$ and azimuthal component $\kappa_\theta$. Clearly, the external thermal field is undisturbed when $\kappa_r \kappa_\theta \sim \kappa_1^2$. By solving the equation (1), $\kappa_1$ can be obtained. As a result, the only target is to concentrate the thermal flux into the core region. According to the previous work, thermal concentrator can be obtained when $\kappa_r > \kappa_\theta$ and the performance is positive to the value of $\kappa_r$. Considering that the background material and the outer layer of the bilayer structure are stainless steel (with thermal conductivity $\kappa_b$ = 15W/m·K, electrical conductivity $\sigma_b$ =1.3 × 10$^6$ S/m) and titanium (with thermal conductivity $\kappa_2$ = 14.6W/m·K, electrical conductivity $\sigma_2$ =2.38 × 10$^6$ S/m), respectively. It is worth mentioning that the titanium is chosen since that it has nearly the same thermal conductivity to the background material (stainless steel). This is because the high $\kappa_2$ would attract the thermal flux lines and confine them within the material, while the low $\kappa_2$ would repel the thermal flux lines and prevent them flowing into the inner layer, both of which would reduce the performance of the concentrator. By setting *a*=1.5mm, *b*=5mm, one can obtain *c*=8.65mm to obtain electric current cloak. According to the analysis above, the thermal conductivity of the inner layer of the bilayer structure can be determined as $\kappa_1$=15.03 W/m·K, which is approximate to the one of background material, indicating that the radial component $\kappa_r$ and azimuthal component $\kappa_\theta$ for the inner layer should fulfill $\kappa_r \kappa_\theta \sim 225$ to obtain thermal concentrator. To confirm the theoretical prediction, simulations based on Multiphysics Comsol were carried out and shown in Figure 2a-d. In the simulation, the interior is connected to the ground and 1V potential is applied to the two sides of the domain to generate uniform current. The simulated result for the electric current is provided in Figure 1a and Figure 1c, where the potential distribution and normalized current density are plotted, respectively. It can be seen that the bilayer structure guides the electric current around the interior smoothly and the perturbations only occur in the outer layer of it, while the interior is kept from the current, thus indicating a good electric current cloak performance. Figure 2b and Figure 2d show the simulated temperature profile distribution and normalized heat flux, respectively, where $\kappa_r = 0.1 \kappa_2$ and $\kappa_\theta = 10 \kappa_2$. From those pictures, one can find that the inner layer concentrates the thermal flux into the interior resulting in considerably enhanced energy density (200% enhancement) while keeping the outside one uniform, revealing a good thermal concentrator.

Experimentally, we have fabricated such bilayer structure, which is schematically illustrated in Figure 3a. Here, the inner layer is replaced by fan-shaped structure, which is composed 12 air wedges (with thermal conductivity $\kappa_b = 0.02$ W/m·K, electrical conductivity $\sigma_b = 0$ S/m) and 12 aluminium nitride (AlN) wedges (with thermal conductivity $\kappa_b = 170$ W/m·K, electrical conductivity $\sigma_b = 0$ S/m). The sample is further cut into spindle shape (see in Figure 3b) to generate a uniform electric current in the observed region, which is confirmed by the simulation when 1V point source was applied to the two sides of it. Figure 3c shows the case for the bilayer structure, where the interior is protected from the external field and the external field keeps undistorted as if nothing happens, indicating a good *dc* electric field cloak. Simulation was also carried out to evaluate the thermal concentrator performance (see Figure 3d). It can be found that the temperature gradient in the interior is remarkably enhanced and the external field keeps nearly undisturbed, revealing a predicted thermal concentrator.

Experimentally, the performance of the electric current cloak can be evaluated by measuring the potential distribution along the lines $x=-10$mm and $x=10$mm, as shown in Figure 4. In the measurement, a DC power supply (DH1715A-5) with 10V was applied to the two sides of the sample and the potential distribution was measured using a Multimeter (Agilent 34410A, 6, 1/2Digit Multimeter). Since the resistance of the sample is small, a 10Ω resistance was used for voltage dividing to protect the power supply. The simulation results and measured results (normalized potential distribution) are shown in Figure 4a-b, respectively. As for ideal case, namely homogeneous background, the equipotential lines are straight. As for the object without cloak, the equipotential lines are seriously distorted. When the bilayer structure is adopted, the equipotential lines return to be straight. Good agreement between the simulated and measured results has been obtained, revealing the capacity of electric cloak for such bilayer structure. To character the thermal property of such structure, infrared heat camera (Fluke Ti300) was utilized to measure the temperature profiles. It should be noted that to reduce the heat conduction/convection by air and the high reflection of stainless steel and titanium for the operating wavelengths of the thermal heat camera, 100 um thickness electrical insulation tape with high emissivity beyond 93% was attached to the upper surface of the sample. In the measurement, two sides of the sample were connected to two temperature sources, namely hot water with temperature 80 ° and ice water mixture with temperature 0 °. For comparison, the background material (stainless) without any structure was also measured. Figure 5 shows the measured steady temperature profile. As plotted in Figure 5a, nearly uniform temperature gradient is generated from the higher temperature to lower one, which can be determined as 2 °/mm. As for the bilayer structure (Figure 5b), the thermal flux flows into the interior resulting in a 170% enhancement of temperature gradient (5.4 °/mm), while keeping the outside temperature field nearly undistorted, thus indicating a good thermal concentrator.

## 3. Thermal Cloak and *dc* Electric Concentrator

At this point, we have demonstrated that this bilayer structure can be utilized to realize a device with the capacity of thermal concentrator and electric cloak. It is reasonable to wonder whether it can also be employed to simultaneously achieve thermal cloak and electric concentrator, as shown in Figure 6a-b. Here, we start with the physical model, which is the same to the one in Figure 1a, in which, however, the inner layer is set as a thermal insulator ($\kappa=0$) and electric conductor. To obtain a thermal cloak, the required thermal conductivity for the outer layer should fulfill:

$$\frac{c^2}{b^2} = \frac{\kappa_2 + \kappa_b}{\kappa_2 - \kappa_b} \tag{4}$$

Here, germanium (with thermal conductivity $\kappa_b =$ 65W/m·K, electrical conductivity $\sigma_b = 1.0 \times 10^{-1}$ S/m) is employed as background material, while the outer layer is made of silicon (with thermal conductivity $\kappa_2 =$ 150W/m·K, electrical conductivity $\sigma_2 = 1.0 \times 10^{-1}$ S/m). Considering $a=1.5$mm, $b=5$mm, one can obtain $c=7$mm to realize a thermal cloak. According to the following equation:

$$\frac{c^2}{b^2} = \frac{(\sigma_2 - \sigma_1)(\sigma_2 + \sigma_b)}{(\sigma_2 + \sigma_1)(\sigma_2 - \sigma_b)} \tag{5}$$

The electric conductivity for the inner layer should be $\sigma_2 = 1.0 \times 10^{-1}$ S/m to ensure the external electric field is undistorted. To concentrate the electric current into interior, it should be anisotropic and has a radial component $\sigma_r$ and an azimuthal component $\sigma_\theta$ following the relationship: $\sigma_r \sigma_\theta \sim \sigma_2^2$, $\sigma_r > \sigma_\theta$.

Experimentally, we have fabricated such structure, which is schematically illustrated in Figure 7a. Here, the fan-shaped structure was employed to act as the inner layer to obtain anisotropic parameters. It should be noted that it is difficult to find a perfect thermal insulator ($\kappa=0$). Instead, we employ the 12 wedges made of solidified silver conductive epoxy (with poor thermal conductivity $\kappa=1$W/m·K and excellent electric conductivity $\sigma=1.5\times10^5$ S/m) and 12 air wedges (with thermal conductivity κ=0.02W/m·K) to construct the fan-shaped structure. Simulations were performed to evaluate the performance of the designed bilayer structure. Figure 7b-c shows the simulated results (potential distribution and current density) for *dc* current, where the current is concentrated into the interior resulting in a 200% enhancement for current density, while the external field keeps undistorted, revealing a good electric concentrator. Similarly, the performance can be evaluated by measuring the potential distribution along the lines *x*=-10mm, *x*=10mm and *y*=0cm. The simulation and experimental results of potential distribution are presented in Figure 8a-c. As expected, the equipotential lines are straight along the *x*=-10mm, *x*=10mm due to no distortion for the external field (Figure 8a-b). As for *y*=0cm, the potential gradient is considerably enhanced for the bilayer structure (0.075V/mm) compared to the one without bilayer structure (0.025V/mm), revealing a *dc* electric concentrator. Good agreement can be obtained between the simulated and experimental results, confirming the feasibility of our method. The simulated result for temperature profile is shown in Figure 7d, where the external isothermal lines are nearly undistorted and the interior is protected from the external field, indicating a good thermal cloak. The thermal flux is also provided in Figure 7e, where the nearly no thermal flux flows into the interior. Note that the slight distortion for the external isothermal lines results from the use of imperfect thermal insulator. The measured results for homogeneous background material and the one with bilayer structure are provided in Figure 9. The temperature gradient in the interior for the homogeneous background material is nearly 2 ℃/mm, while the one for bilayer structure can be measured as 0.1 ℃/mm (about 5% of the applied temperature gradient). On the other hand, the external field keeps disturbed as if nothing happens, thereby this device behaves as thermal cloak.

**Conclusion**

We propose a bilayer structure to realize the independent manipulation of *dc* electric and thermal fields by directly solving the *dc* electric / thermal field equations. The experimental realization of a structure simultaneously behaving as thermal concentrator (cloak) and *dc* electric cloak (concentrator) confirms the feasibility of our methodology. Since it is not based on TO, it can be easily experimentally achieved with naturally occurring materials and can be readily scaled up and down, indicating that our job provides a novel and practical avenue to manipulation of multiphysics fields, thus might find potential applications in various areas.

This work was supported by the National Natural Science Foundation of China under Grant Nos. 51032003, 11274198, 51221291 and 61275176, National High Technology Research and Development Program of China under Grant No. 2012AA030403, Beijing Municipal Natural Science Program under Grant No. Z141100004214001, and the Science and technology plan of Shenzhen city under grant Nos.JCYJ20120619152711509, JC201105180802A and CXZZ20130322164541915.

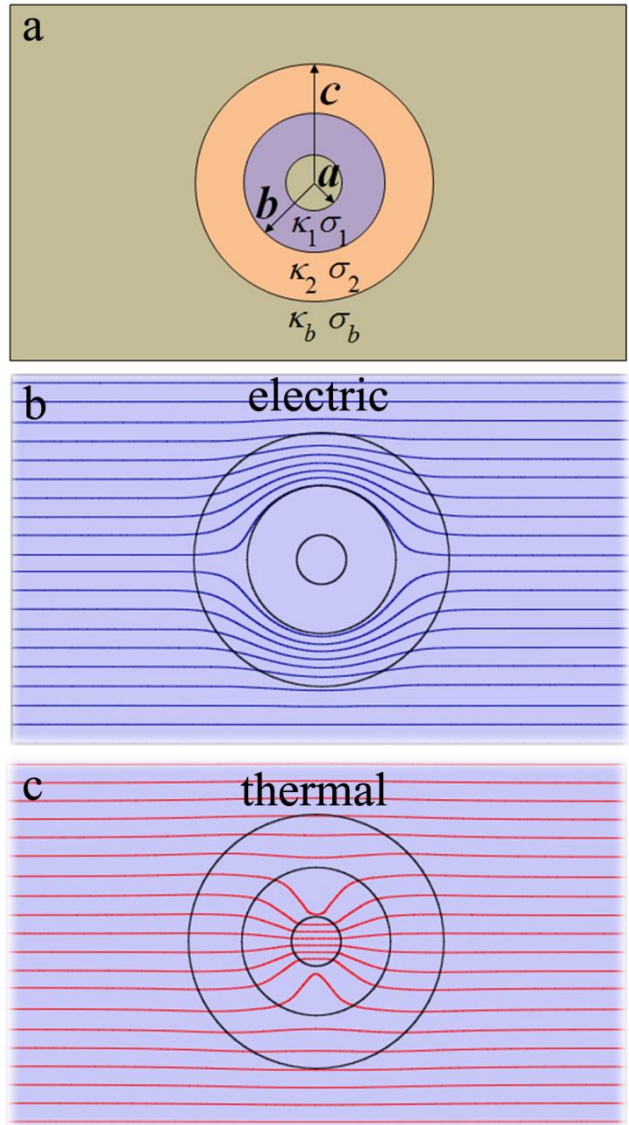

**Figure 1.** The principle for bilayer structure behaving as *dc* electric cloak and thermal concentrator: a) The corresponding physical model. b) The current distribution. c) The thermal flux distribution.

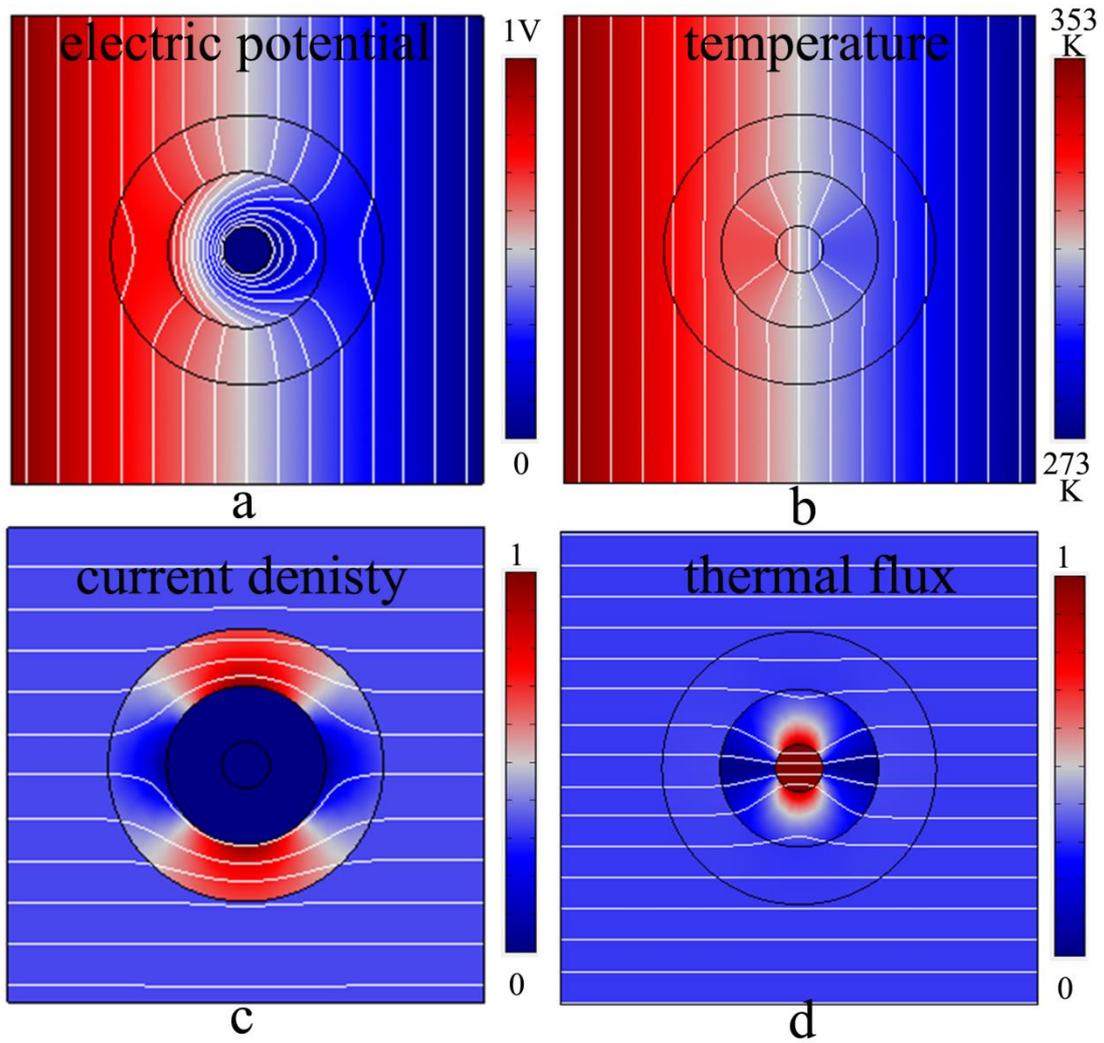

**Figure 2.** The simulated results for the bilayer structure: a) The electric potential distribution. b) The normalized current distribution. c) The simulated temperature profile. d) The normalized thermal flux distribution.

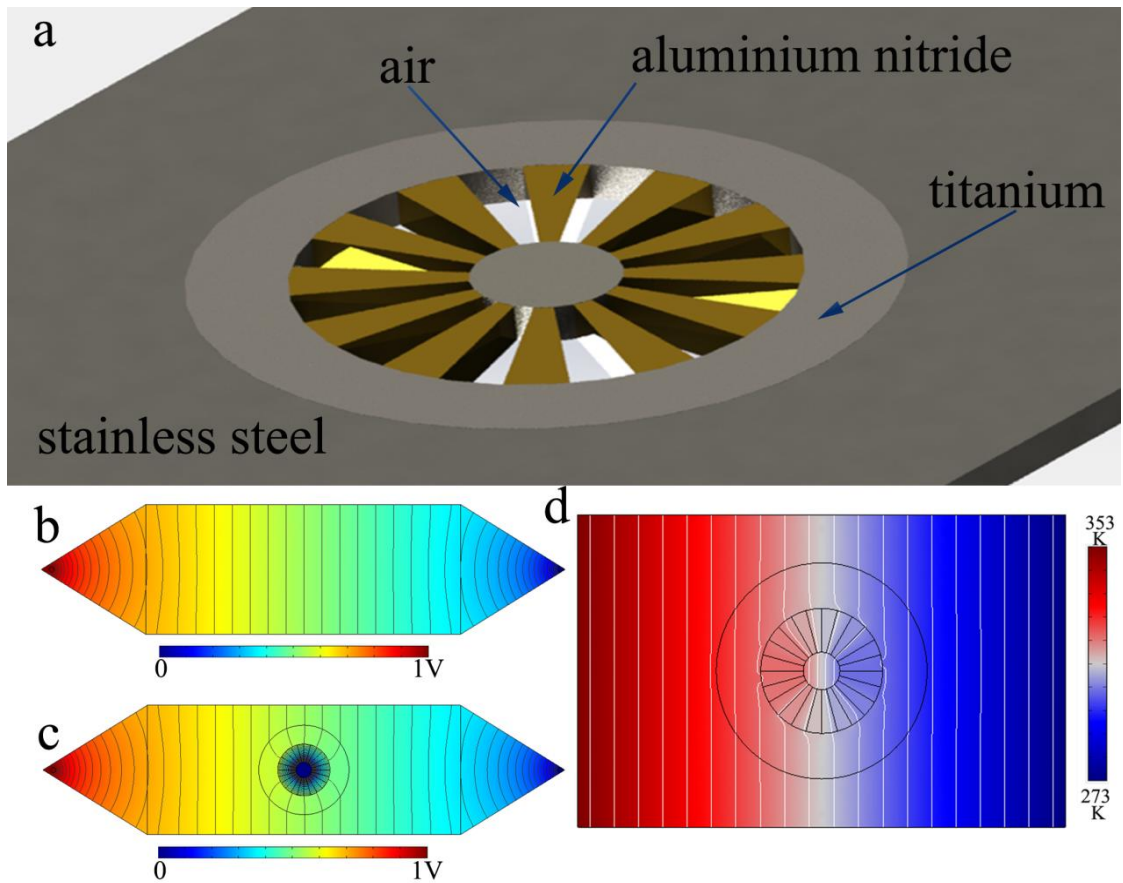

**Figure 3.** (a) The schematic illustration for practical realization of bilayer structure. (b) The simulated potential distribution for a homogeneous spindle shape background medium. The black lines represent the equipotential lines. (c) The simulated potential distribution for a spindle shape background medium with bilayer structure. (d) The simulated temperature profile for bilayer structure. The white lines represent the isotemperature lines.

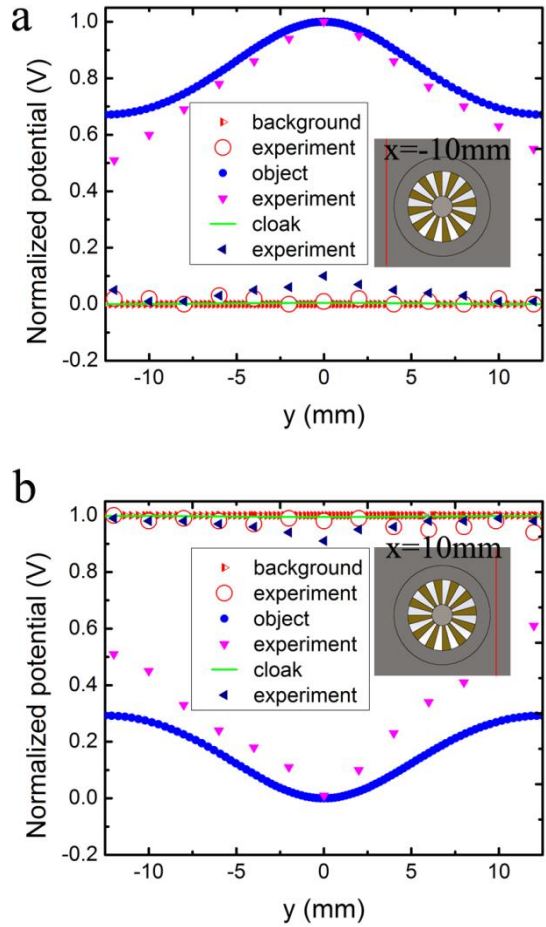

**Figure 4.** Simulation and experimental results of potential distribution for the bilayer structure: (a) The normalized potential distribution at the left observation line, (b) The normalized potential distribution at the right observation line. The observation lines are marked in red.

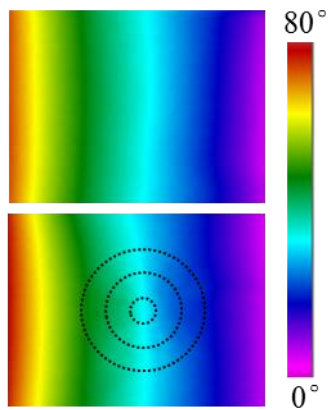

**Figure 5.** Measured temperature profiles: (a) homogeneous background material. (b) with bilayer structure. The black dash circles represent the bilayer structure.

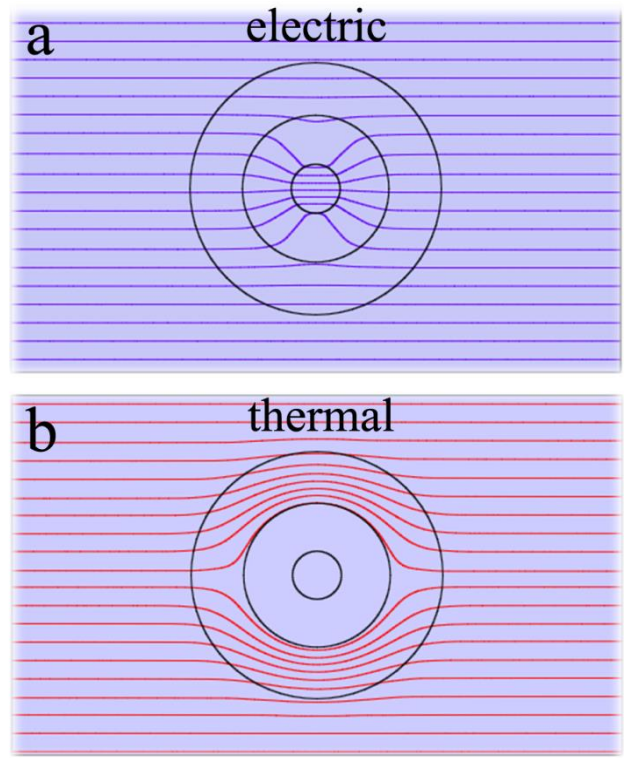

**Figure 6.** The principle for bilayer structure behaving as *dc* electric concentrator and thermal cloak: (b) The current distribution. (c) The thermal flux distribution.

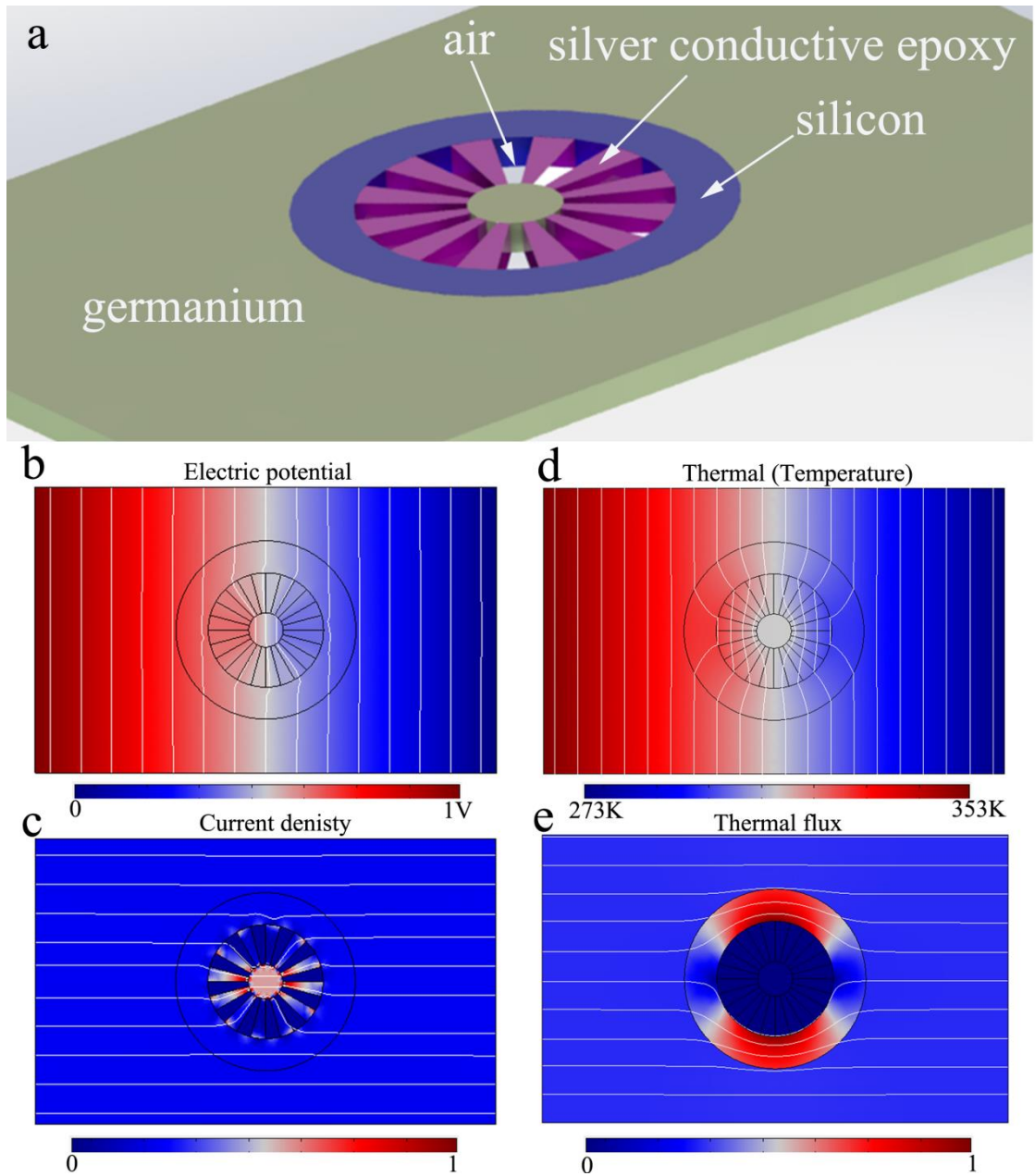

**Figure 7.** (a) The schematic illustration for practical realization of bilayer structure behaving as electric concentrator and thermal cloak. (b) (c) The simulated potential distribution and normalized current density. (d) (f) The simulated temperature profile and normalized thermal flux.

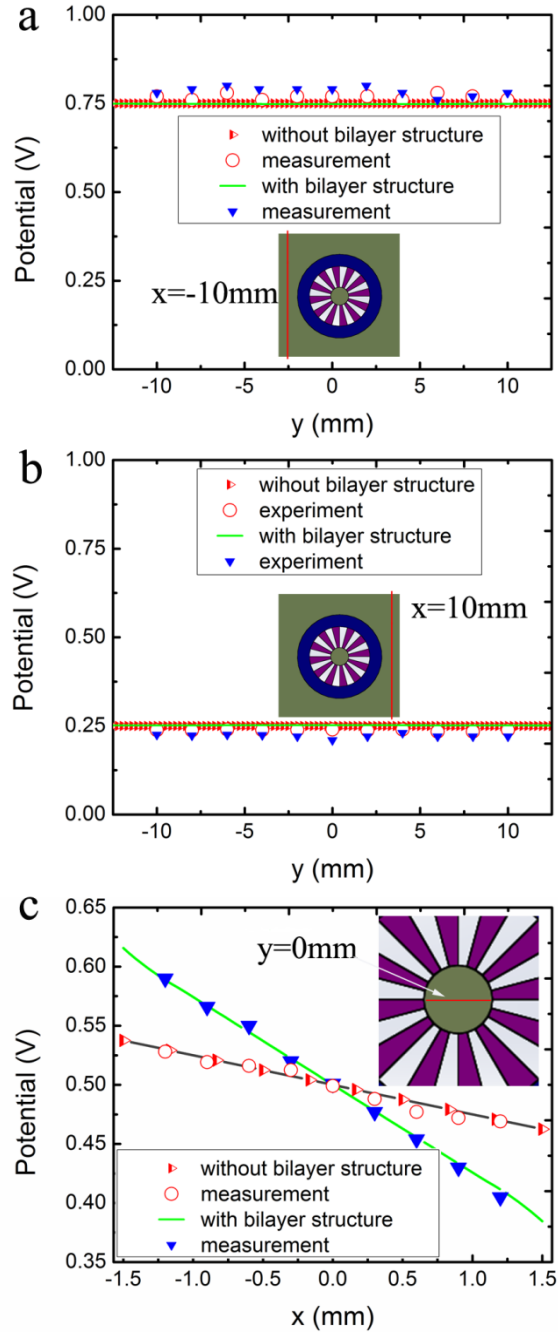

**Figure 8.** Simulation and experimental results of potential distribution for the bilayer structure behaving as electric concentrator: (a) The potential distribution at the left observation line. (b) The potential distribution at the right observation line. (c) The potential distribution at the internal observation line. The observation lines are marked in red.

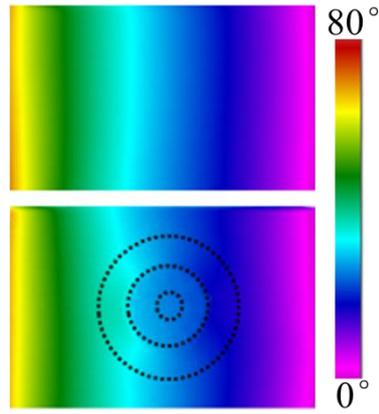

**Figure 9.** Measured temperature profiles: (a) homogeneous background material. (b) with bilayer structure.